\documentclass[aps,prl,twocolumn,showpacs,amsmath,amssymb]{revtex4-1}
\usepackage{epsfig}
\usepackage{graphicx}% Include figure files
\usepackage{dcolumn}% Align table columns on decimal point
\usepackage{bm}% bold math
\usepackage{ltablex,booktabs}
\usepackage{overpic}
\usepackage{subfigure}
\usepackage{float}
\usepackage{color}
\usepackage{amsmath}
\usepackage{mathcomp}
\usepackage{mathrsfs}
\usepackage{multirow}
\usepackage{rotating}
\usepackage{amssymb}
\usepackage{gensymb}
\usepackage{amsmath}
\usepackage{tabularx}
\usepackage[bookmarksnumbered, pdfstartview=FitH,colorlinks,urlcolor=blue, citecolor=blue,linkcolor=blue] {hyperref}
\begin{document}
\normalsize
\parskip=5pt plus 1pt minus 1pt

%\setpagewiselinenumbers
%\linenumbers
\title{\boldmath Observation of $a_0(1710)^+ \to K_S^0K^+$ in study of the $D_s^+\to K_S^0K^+\pi^0$ decay}
%\vspace{-1cm}
%\author{Author list}
\author{
\begin{small}
\begin{center}
M.~Ablikim$^{1}$, M.~N.~Achasov$^{10,b}$, P.~Adlarson$^{69}$, M.~Albrecht$^{4}$, R.~Aliberti$^{29}$, A.~Amoroso$^{68A,68C}$, M.~R.~An$^{33}$, Q.~An$^{65,51}$, X.~H.~Bai$^{59}$, Y.~Bai$^{50}$
% ...
\\
\vspace{0.2cm}
(BESIII Collaboration)\\
\vspace{0.2cm} {\it
$^{1}$ Institute of High Energy Physics, Beijing 100049, People's Republic of China\\
% ...
$^{4}$ Bochum Ruhr-University, D-44780 Bochum, Germany\\
% ...
$^{10}$ G.I. Budker Institute of Nuclear Physics SB RAS (BINP), Novosibirsk 630090, Russia\\
% ...
$^{29}$ Johannes Gutenberg University of Mainz, Johann-Joachim-Becher-Weg 45, D-55099 Mainz, Germany\\
% ...
$^{33}$ Liaoning Normal University, Dalian 116029, People's Republic of China\\
% ...
$^{50}$ Southeast University, Nanjing 211100, People's Republic of China\\
$^{51}$ State Key Laboratory of Particle Detection and Electronics, Beijing 100049, Hefei 230026, People's Republic of China\\
% ...
$^{59}$ University of Jinan, Jinan 250022, People's Republic of China\\
% ...
$^{65}$ University of Science and Technology of China, Hefei 230026, People's Republic of China\\
% ...
$^{68}$ University of Turin and INFN, (A)University of Turin, I-10125, Turin, Italy; (B)University of Eastern Piedmont, I-15121, Alessandria, Italy; (C)INFN, I-10125, Turin, Italy\\
$^{69}$ Uppsala University, Box 516, SE-75120 Uppsala, Sweden\\
% ...
$^{b}$ Also at the Novosibirsk State University, Novosibirsk, 630090, Russia\\
% ...
}
\end{center}
\vspace{0.4cm}
\end{small}
}
\noaffiliation{}

%\author{\input{author}}
%\affiliation{}
%\vspace{-10cm}
%\date{\today}
%\setpagewiselinenumbers
%------------------------------------------------------------------------------
\begin{abstract}
  Using $e^+e^-$ annihilation data corresponding to an integrated luminosity of
  6.32 fb$^{-1}$ collected at center-of-mass energies between 4.178~GeV and
  4.226~GeV with the BESIII detector, we perform the first amplitude analysis
  of the decay $D_s^+\to K_S^0K^+\pi^0$ and determine the relative branching
  fractions and phases for intermediate processes. We observe the
  $a_0(1710)^+$, the isovector partner of the $f_0(1710)$ and $f_0(1770)$
  mesons, in its decay to $K_S^0K^+$ for the first time. In addition, we
  measure the ratio 
  $\frac{\mathcal{B}(D_{s}^{+} \to \bar{K}^{*}(892)^{0}K^{+})}{\mathcal{B}(D_{s}^{+} \to \bar{K}^{0}K^{*}(892)^{+})}$
  to be $2.35^{+0.42}_{-0.23\text{stat.}}\pm 0.10_{\rm syst.}$. Finally, we
  provide a precision measurement of the absolute branching fraction
  $\mathcal{B}(D_s^+\to K_S^0K^+\pi^0) = (1.46\pm 0.06_{\text{stat.}}\pm 0.05_{\text{syst.}})\%$.
\end{abstract}
%\pacs{}
\maketitle

%------------------------------------------------------------------------------
%\section{Introduction}
%------------------------------------------------------------------------------
The constituent quark model describes mesons as bound $q\bar{q}$ states grouped
into SU(3) flavor multiplets. In this scenario, the $f_0(500)$ and $f_0(980)$
are often classified as the ground states of the isoscalar scalar mesons and 
the $a_0(980)$ meson is taken as their isovector partner. The $f_0(1370)$,
$f_0(1500)$, and $a_0(1450)$ are then considered to be the corresponding
radially excited states. Within the next higher set of excitations, however,
which includes the $f_0(1710)$ and $f_0(1770)$, the isovector scalar
meson~(i.e.~the $a_0(1710)$) has been proposed but has not yet been well
established~\cite{Oset1, Oset2, Klempt, plb-816-136227}. The $f_0(1710)$ is
often considered to be a likely candidate for a glueball or $K^*K^*$ molecule.
Although the recent measurement of the branching fraction~(BF) ratio 
$\mathcal{B}(f_0(1710) \to \eta \eta^{\prime})/\mathcal{B}(f_0(1710)\to \eta \eta^{\prime})$~\cite{ref:arXiv2202.00621, ref:arXiv2202.00623}
supports the hypothesis that the $f_0(1710)$ has a large glueball compenent,
one decisive way to determine whether the $f_0(1710)$ is a glueball or a
$K^*K^*$ molecular is to search for an isovector partner, the $a_0(1710)$.
The $a_0(1710)^{+}$ meson was previously reported by the BaBar experiment in
the process $\eta_c \to a_0(1710)^{+} \pi^-$ with
$a_0(1710)^{+}\to \pi^{+}\eta$~\cite{a01710}. In addition, the BESIII
experiment reported interference between the $f_0(1710)$ and $a_0(1710)^{0}$
in amplitude analyses of $D_s^+\to K_S^0K_S^0\pi^+$ and
$D_s^+\to K^+K^-\pi^+$~\cite{KSKSpi, MWang}. However, more studies of $D_s^+$
meson decays are crucial to firmly establish the $a_0(1710)$ triplet.
Ref.~\cite{Oset2} predicts the product BF of $D_s^+\to a_0(1710)^{+} \pi^0$
with $a_0(1710)^{+} \to K_S^0K^+$ to be $(2.0\pm0.7)\times 10^{-3}$. An
amplitude analysis of $D_s^+\to K_S^0K^+\pi^0$ therefore provides an ideal
opportunity to study the $a_0(1710)^{+} \to K_S^0K^+$ decay.

%------------------------------------------------------------------------------
The internal quark structure of the light scalar mesons, like the $a_0(980)$,
have also been the source of much theoretical speculation.  They have been
considered to be~$q\bar{q}$, $qq\bar{q}\bar{q}$, $K\bar{K}$, etc. The coupling
constants, $g_{a_0\pi\eta}$ and $g_{a_0KK}$, are predicted by various
models~\cite{NPB-315-465, PRD-56-4084, PLB-759-501} and therefore serve as
important experimental constraints on theoretical models. Combining an analysis
of $D_s^+\to K_S^0K^+\pi^0$ with a previous measurement of
$D_{s}^{+} \to \pi^+\pi^0\eta$~\cite{Doc-DB-682-v7},
we can determine the ratio
$\frac{\mathcal{B}(a_0(980)\to K\bar{K})}{\mathcal{B}(a_0(980)\to \eta\pi)}$.
This is a key experimental input for the calculation of the coupling constants
of the $a_0(980)$ and helps determine its quark
composition~\cite{NPB-315-465, PRD-56-4084, PLB-759-501, prd-82-034016, BCKa0, BCKa02}.

%------------------------------------------------------------------------------
Furthermore, Ref.~\cite{PRD-100-093002} predicts that
$\mathcal{B}(D_{s}^{+} \to \bar{K}^{*}(892)^{0}K^+)$ is larger than
$\mathcal{B}(D_{s}^{+} \to \bar{K}^{0}K^{*}(892)^{+})$, but the current
experimental uncertainties are too large to verify this~\cite{PDG}. In an
analysis of $D_s^+\to K_S^0K^+\pi^0$, we can measure the BFs of both modes
simultaneously. Thus, the correlated systematic uncertainties arising from the
masses and widths of the resonances, the model parameters, and the common
backgrounds can be considered and reduced.

Because of its large BF, the Cabibbo-favored $D_{s}^{+} \to K_{S}^{0}K^+\pi^{0}$
decay is one of the golden decay channels of the $D_{s}^{+}$. This decay can be
used as a reference channel for other decays of the $D^+_{s}$ meson and it is
important for our understanding of $B^0_s$ decays to final states involving the
$D_{s}^{+}$ mesons~\cite{PDG}. The CLEO experiment measured the absolute BF of
the $D_{s}^{+} \to K_{S}^{0}K^+\pi^{0}$ decay to be
$(1.52\pm 0.09_{\rm stat.}\pm0.20_{\rm syst.})\%$~\cite{CLEO-BF}, using 586
pb$^{-1}$ of $e^+e^-$ collisions recorded at a center-of-mass energy of
4.17~GeV. 

In this Letter, we present the first amplitude analysis and a more precise
measurement of the BF for the decay $D_{s}^{+} \to K_{S}^{0}K^+\pi^{0}$ using
6.32~$\rm fb^{-1}$ of data collected with the BESIII detector at center-of-mass
energies between 4.178 and 4.226~GeV. Charge-conjugated modes are implied
throughout this paper.
%------------------------------------------------------------------------------

The BESIII detector~\cite{Ablikim:2009aa, Ablikim:2019hff} records symmetric
$e^+e^-$ collisions provided by the BEPCII storage
ring~\cite{Yu:IPAC2016-TUYA01}. The cylindrical core of the BESIII detector
covers 93\% of the full solid angle and consists of a helium-based multilayer
drift chamber~(MDC), a plastic scintillator time-of-flight system~(TOF), and a
CsI(Tl) electromagnetic calorimeter~(EMC), which are all enclosed in a
superconducting solenoidal magnet providing a 1.0~T magnetic field. The end cap
TOF system was upgraded in 2015 using multi-gap resistive plate chamber
technology~\cite{etof}.

%------------------------------------------------------------------------------
Simulated data samples produced with a {\sc geant4}-based~\cite{geant4} Monte 
Carlo (MC) package, which includes the geometric description of the BESIII 
detector and the detector response, are used to determine detection
efficiencies and to estimate backgrounds. The beam energy spread and initial 
state radiation (ISR) in the $e^+e^-$ annihilations are simulated with the
generator {\sc kkmc}~\cite{ref:kkmc}. The inclusive MC sample includes the
production of open charm processes, the ISR production of vector
charmonium(-like) states, and the continuum processes incorporated in
{\sc kkmc}~\cite{ref:kkmc}. The known decay modes are described with
{\sc evtgen}~\cite{ref:evtgen} using BFs taken from the Particle Data
Group~\cite{PDG}, and the remaining unknown charmonium decays are generated
with {\sc lundcharm}~\cite{ref:lundcharm}. Final state radiation~(FSR) from
charged final state particles is incorporated using {\sc photos}~\cite{photos}.

%------------------------------------------------------------------------------
We reconstruct the process 
$e^{+}e^{-} \to D_{s}^{*+}D_{s}^{-}\to \gamma D_{s}^{+}D_{s}^{-}$ using both
single-tag~(ST) and double-tag~(DT) candidate events~\cite{MarkIII-tag}. An ST
candidate is an event where only the $D_{s}^{-}$ meson is reconstructed through
particular hadronic decays (tag modes) without any requirement on the remaining
measured tracks and EMC showers. A DT candidate is an event where the
$D_{s}^{+}$ is reconstructed through $D_{s}^{+} \to K^0_{S}K^+\pi^{0}$ in
addition to the $D_{s}^{-}$ being reconstructed through the tag modes. Eight
tag modes are used:
$D_s^-\to K_{S}^{0}K^{-}$, $K^{+}K^{-}\pi^{-}$, $K^{+}K^{-}\pi^{-}\pi^{0}$,
$K_{S}^{0}K^{-}\pi^{-}\pi^{+}$, $K_{S}^{0}K^{+}\pi^{-}\pi^{-}$,
$\pi^{-}\pi^{-}\pi^{+}$, $\pi^{-}\eta^{\prime}$, and $K^{-}\pi^{-}\pi^{+}$. 
The selection criteria for the final state particles, transition photon, and
the $D_{s}^{\pm}$ candidates are the same as
Refs.~\cite{ref:a0980, ref:Kspipi0, ref:KsKpipi}. The $K_{S}^{0}$, $\pi^{0}$,
$\eta$ and $\eta^{\prime}$ mesons are reconstructed through
$K_{S}^{0} \to \pi^{+}\pi^{-}$, $\pi^{0} \to \gamma\gamma$,
$\eta \to \gamma\gamma$ and $\eta^{\prime} \to  \pi^{+}\pi^{-}\eta$ decays,
respectively. 

%------------------------------------------------------------------------------
An eight-constraint kinematic fit is applied to the DT candidates to select
signal events for the amplitude analysis. The total four-momentum is
constrained to the four-momentum of the $e^+e^-$ system, and the invariant
masses of the $K_S^0$, $\pi^0$, $D_s^-$, and $D_s^{*+(-)}$ candidates are
constrained to their corresponding known masses~\cite{PDG}. Within each event,
the candidate with the minimum $\chi^2$ from the kinematic fit is chosen. The
invariant mass of the signal $D_s^+$ is then required to be within (1.930,
1.990)~GeV/$c^2$. A ninth constraint, on the mass of the signal $D_s^+$, is
then added to the kinematic fit to guarantee all candidates lie inside the
allowed phase-space. There are 1050 DT events obtained for the amplitude
analysis with a signal purity of~($94.7\pm0.7$)\%, which is determined from a
fit to the invariant mass distribution of the signal $D_s^+$ candidates.
%------------------------------------------------------------------------------

The intermediate resonance composition is determined using an unbinned
maximum-likelihood fit. The likelihood function is described by a signal
probability density function~(PDF), $\left|\mathcal{M}(p_{j})\right|^{2}$,
incoherently added to a background PDF, denoted as
$B$~\cite{ref:Kspipi0, ref:KsKpipi, prd-85-122002}. The signal amplitude
$\mathcal{M}$ is constructed based on the isobar model
formulation~\cite{covariant-tensors}. The background PDF is constructed from
inclusive MC samples using RooNDKeysPdf~\cite{Verkerke}. RooNDKeysPdf is a
kernel estimation method~\cite{Cranmer} implemented in RooFit~\cite{Verkerke},
which models the distribution of an input dataset as a superposition of
Gaussian kernels. The likelihood function is then written as
\begin{eqnarray}
  \begin{aligned}
    \mathcal{L}=\prod_{k} &\left[\frac{\epsilon f_s\left|\mathcal{M}(p_{k}^{\mu})\right|^{2}\,R_3}{\int \epsilon\left|\mathcal{M}(p_{k}^{\mu})\right|^{2}\,R_{3}dp_k}
      +\frac{(1-f_s)B(p_{k}^{\mu})R_{3}}{\int \epsilon B(p_{k}^{\mu})\,R_{3}dp_k}\right], \label{combined-PDF-2}
  \end{aligned}
\end{eqnarray}
where $\epsilon$ is the acceptance function, the index $k$ runs over selected
events, $p_{k}^{\mu}$ represents the four-momenta of the final particles in the
$k^{\rm th}$ event, $f_s$ is the signal purity, and $R_{3}$ is an element of
three-body phase space. The normalization integral in the denominator is
calculated by MC integration~\cite{ref:Kspipi0}.

The signal amplitude $\mathcal{M}$ is a coherent sum of the amplitudes for the
intermediate processes,
$\mathcal{M}=\begin{matrix}\sum c_{n}\mathcal{A}_{n}\end{matrix}$, where $n$
indicates the $n^{\rm th}$ intermediate state. The complex coefficient $c_{n}$
equals $\rho_{n}e^{i\phi_{n}}$ with magnitude $\rho_{n}$ and phase $\phi_{n}$.
The amplitude $\mathcal{A}_{n}$ is the product of the spin
factor~\cite{covariant-tensors}, the Blatt-Weisskopf barriers of the
intermediate state and the $D_{s}^{+}$ meson~\cite{Blatt}, and the relativistic
Breit-Wigner function~\cite{RBW} to describe the propagator for the
intermediate resonance.

\begin{figure}[!htp]
  \centering
  \includegraphics[width=0.235\textwidth]{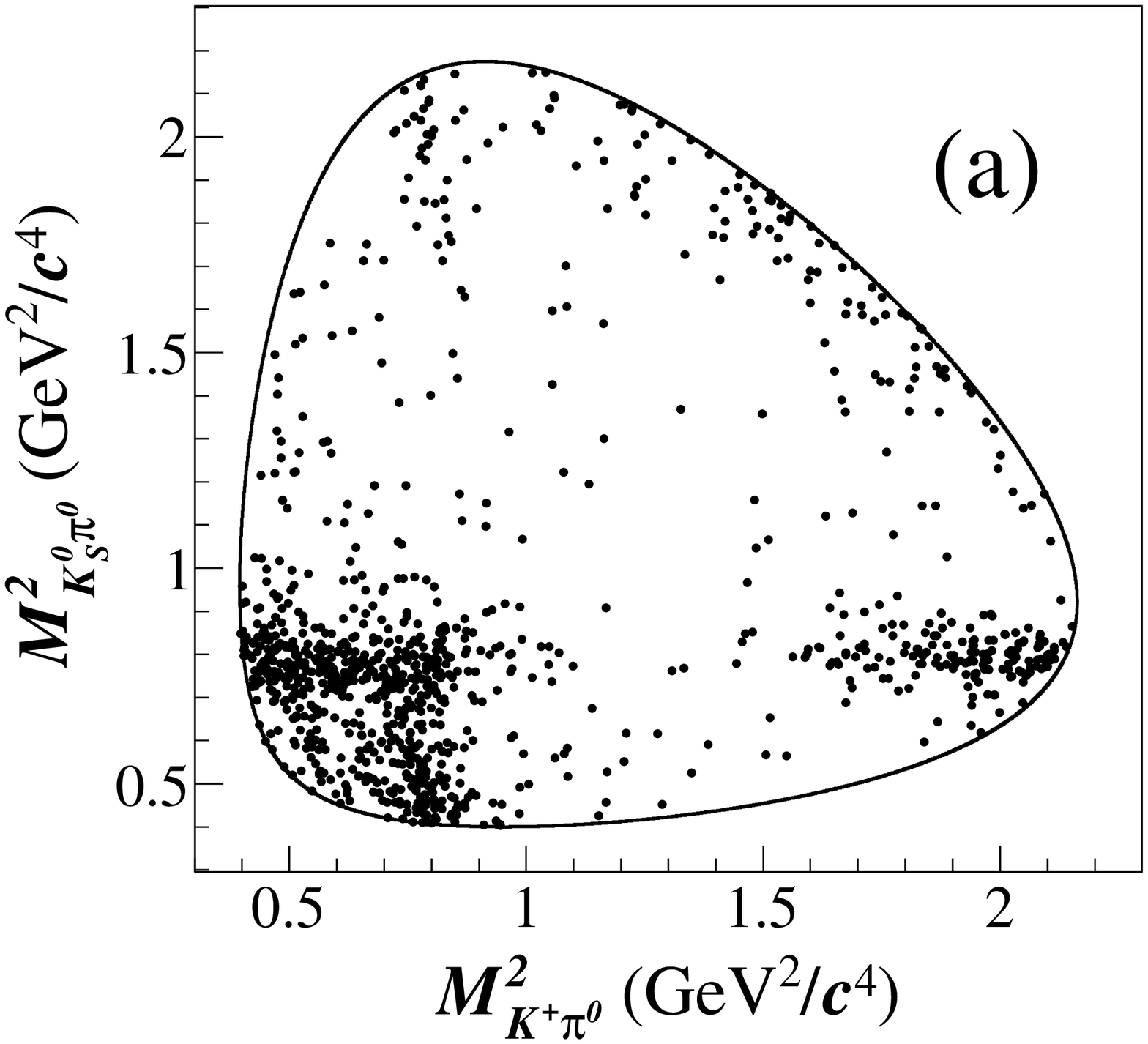}
  \includegraphics[width=0.235\textwidth]{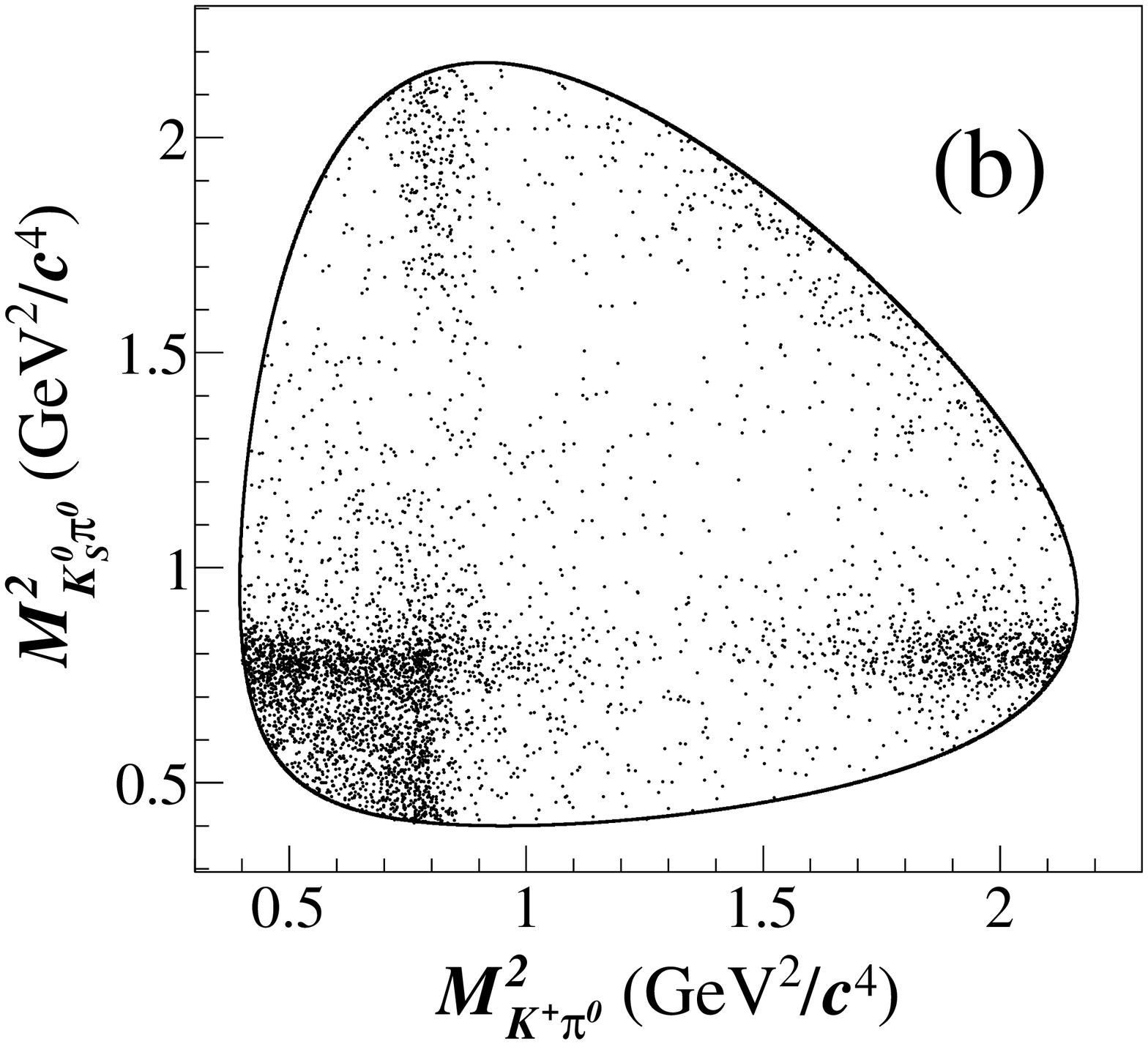}
  \caption{The Dalitz plot of $M^{2}_{K_{S}^{0}\pi^{0}}$
    versus~$M^{2}_{K^{+}\pi^{0}}$ for the decay
    $D^{+}_{s}\to K^{0}_{S}K^{+}\pi^{0}$ from (a) the data sample and (b) the
    inclusive MC sample generated based on the results of the amplitude
    analysis. The black curve indicates the kinematic boundary.}
  \label{fig:dalitz}
\end{figure}

%------------------------------------------------------------------------------
The $M^2_{K_S^0\pi^0}$ versus $M^2_{K^+\pi^0}$ Dalitz plot, shown in
Fig.~\ref{fig:dalitz}, reveals there is a strong contribution from the process
$D_s^+\to \bar{K}^{*}(892)^0K^+$, which appears as the horizontal band around
0.8~GeV$^2$/$c^4$.
Besides this dominant intermediate process, other possible intermediate
resonances are considered, including the $K^*_0(700)$, $K^*(892)$, $K^*(1410)$,
$K_0^*(1430)$, $K_2^*(1430)$, $K^*(1680)$, $a_{0}(980)$, $a_{2}(1320)$,
$a_{0}(1450)$, $a_{2}(1700)$, $a_{0}(1710)$, $\rho(1700)$, and the
$(K\pi)_{\rm S-wave}$~(using the LASS parameterization~\cite{KpiLASS} and the
K-matrix~\cite{Kmatrix}).  Each possibility is added to the fit one at a time.
Various combinations of these resonances are tested as well. The statistical
significance of each amplitude is calculated based on the change of the
log-likelihood with and without this amplitude after taking the change of the
degrees of freedom into account. If the significance of a newly added amplitude
is larger than 5$\sigma$, this amplitude is kept, otherwise it is dropped.
During the fit, $f_s$ is fixed and the magnitudes and phases of all
intermediate processes are floating and are measured with respect to those of
the $D_s^+\to \bar{K}^{*}(892)^0K^+$. The mass and width of the $a_{0}(1710)^+$
are free, those of the $a_{0}(980)^+$ are fixed to the values given in
Ref.~\cite{980mw}, and those of all other resonances are fixed to their known
values~\cite{PDG}. 
%------------------------------------------------------------------------------

\begin{figure*}[!htbp]
  \centering
  \includegraphics[width=0.3\textwidth]{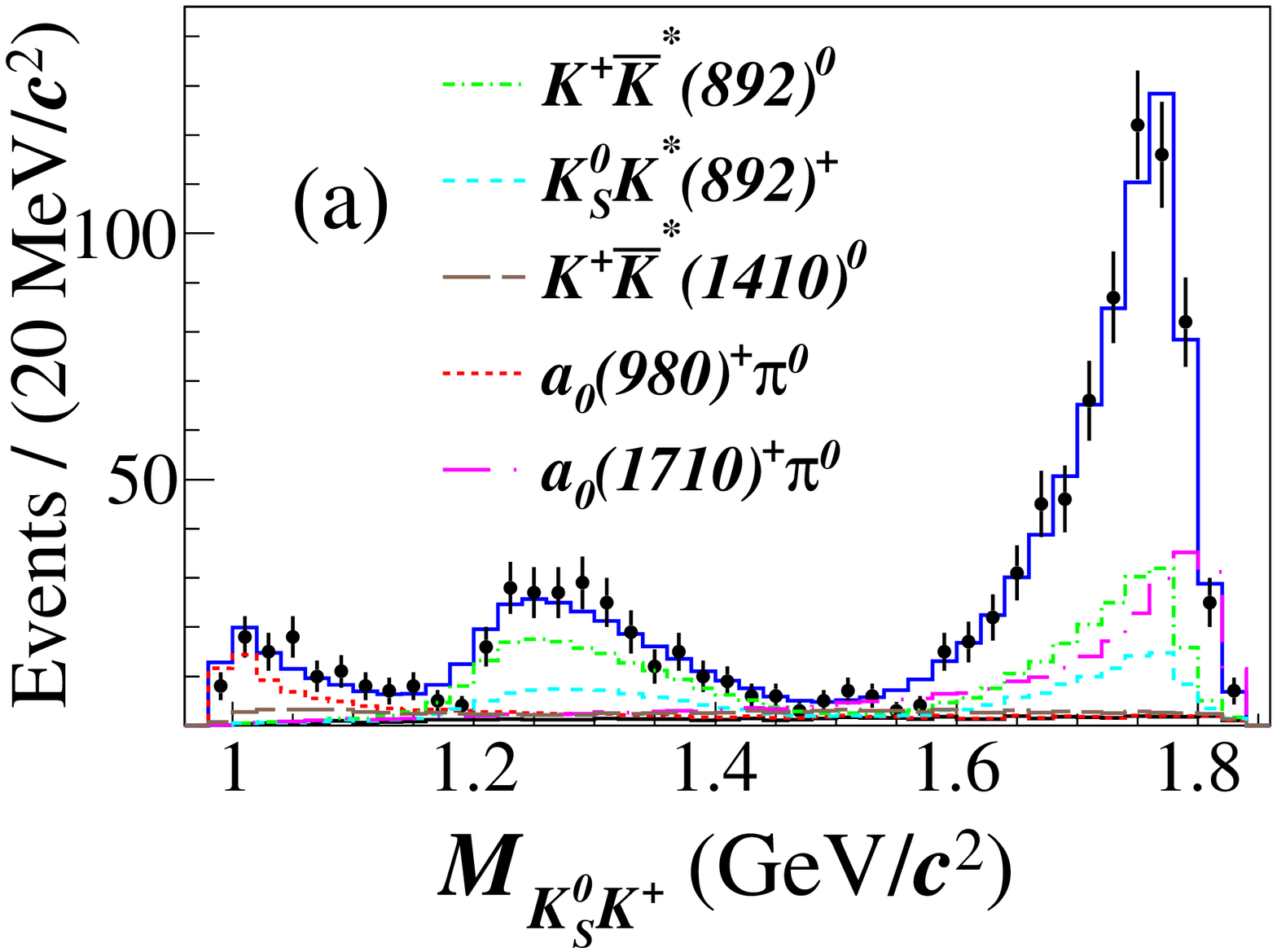}
  \includegraphics[width=0.3\textwidth]{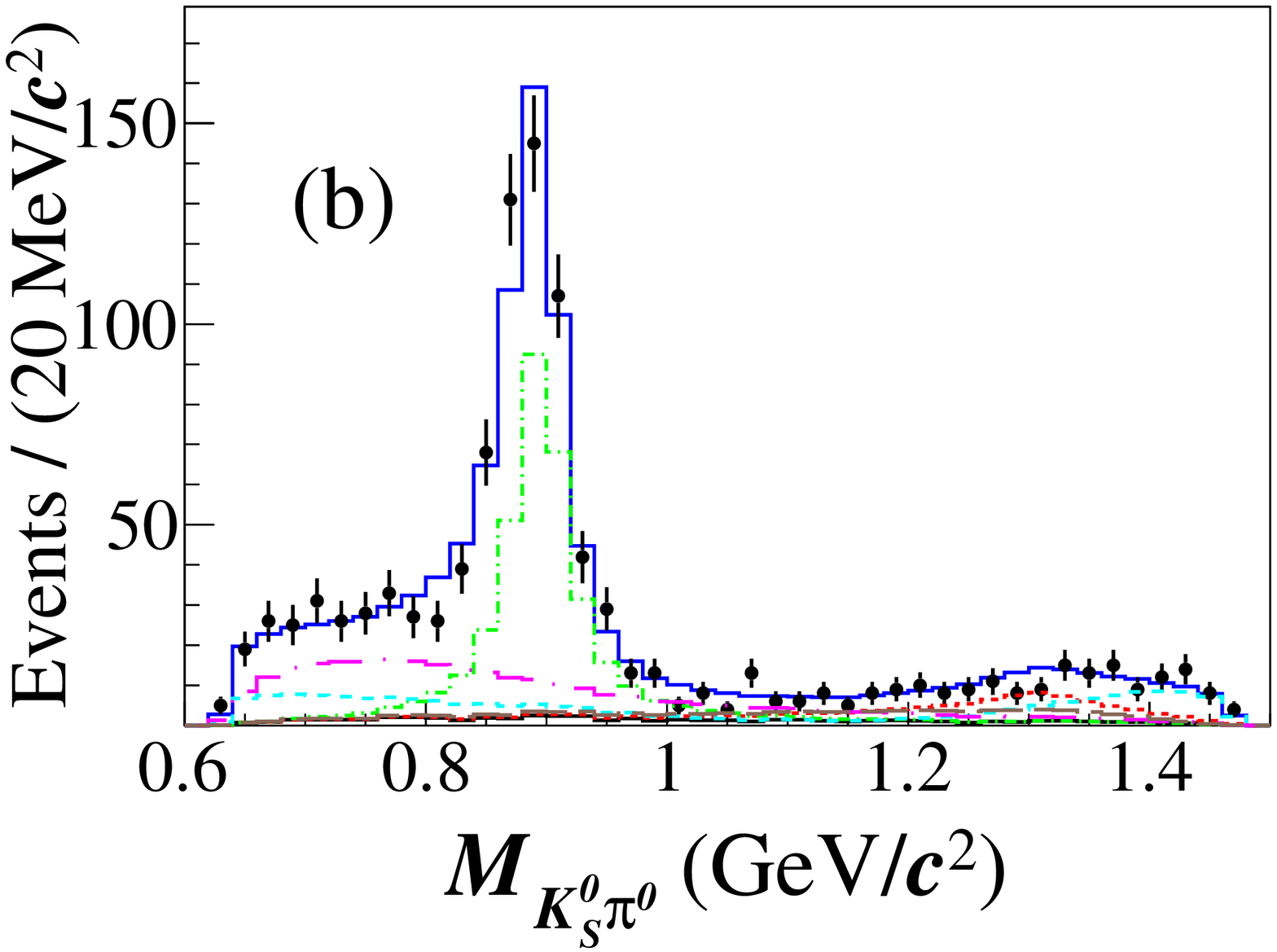}
  \includegraphics[width=0.3\textwidth]{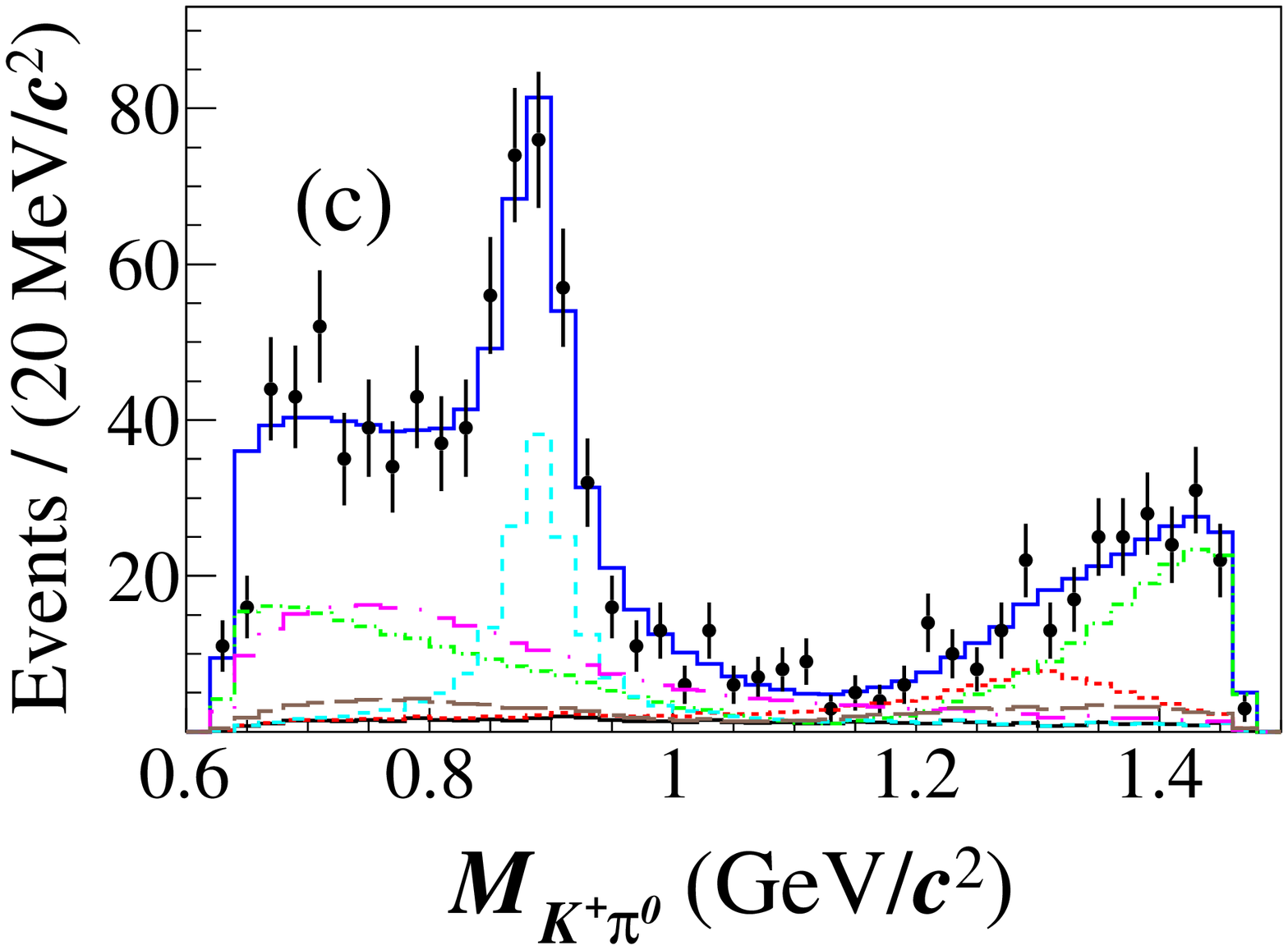}
  \caption{
    The projections of the Dalitz plot onto (a) $M_{K_S^0K^+}$, (b)
    $M_{K_S^0\pi^0}$, and (c) $M_{K^+\pi^0}$. The data samples are represented
    by points with error bars, the fit results by blue lines, and backgrounds
    by black lines. Colored dashed lines show the components of the fit model.
    Due to interference effects, the total of the FFs is not necessarily equal
    to the sum of the components.}
  \label{dalitz-projection}
\end{figure*}

Five intermediate processes, $D_s^+ \to \bar{K}^{*}(892)^0K^+$,
$D_s^+\to K^{*}(892)^+K_S^0$, $D_s^+\to a_{0}(980)^+\pi^0$,
$D_s^+\to \bar{K}^*(1410)^{0}K^+$, and $D_s^+\to a_{0}(1710)^+\pi^0$, are
eventually retained as the optimal set. The mass projections of the fit result
are shown in Fig.~\ref{dalitz-projection}. The contribution of the $n$th
intermediate process relative to the total BF is quantified by a fit
fraction~(FF) defined as
${\rm FF}_{n} = \int \left|\rho_{n}\mathcal{A}_{n}\right|^{2}R_3dp_j/\int\left|\mathcal{M}\right|^{2}R_3dp_j$.
The ratio
$\frac{\mathcal{B}(D_{s}^{+} \to \bar{K}^{*}(892)^{0}K^{+})}{\mathcal{B}(D_{s}^{+} \to \bar{K}^{0}K^{*}(892)^{+})}=2.35^{+0.42}_{-0.23\text{stat.}}\pm 0.10_{\rm syst.}$
is calculated as the quotient of their FFs, where correlations are accounted
for in the systematic and statistical uncertainties. The phases and FFs for the
intermediate processes are listed in Table~\ref{fit-result}. The mass and width
of the $a_0(1710)^+$ are 
($1.817\pm0.008_{\rm stat.}\pm0.020_{\rm syst.}$)~GeV/$c^{2}$ and
($0.097\pm 0.022_{\rm stat.} \pm 0.015_{\rm syst.}$)~GeV/$c^{2}$, respectively.
%------------------------------------------------------------------------------

Some tests are made to further clarify the existence of the $a_0(1710)^+$.
First, the recoil of the $K^*_0(700)$ may cause an enhancement at the high end
of the $K_S^0K^+$ spectrum, but the shape of the $K^*_0(700)$ does not match
the distribution of data and has a significance less than $3\sigma$. Second,
the log-likelihood value of a fit with the $K^*_0(700)$ included and the
$a_0(1710)^+$ excluded decreases by 40 compared to the nominal fit. In
addition, even though the $\rho(1700)^+$ and the $a_2(1700)^+$ peak at the
same position as the $a_0(1710)^+$ in the $K_S^0K^+$ spectrum, the
log-likelihood value is worse by 70 units when these resonances are included
instead of the $a_0(1710)^+$. 
%------------------------------------------------------------------------------
%------------------------------------------------------------------------------
\begin{table*}[!htbp]
  \caption{Phases, FFs, BFs, and statistical significances~($\sigma$) of
    intermediate processes with the final state $K_S^0K^+\pi^0$. The first and
    second uncertainties are statistical and systematic, respectively. Due to
    interference effects, the total of the FFs is not necessarily equal to the
    sum of the components.}
  \label{fit-result}
  \begin{center}
    \begin{tabular}{lccccc}
      \hline \hline
      Amplitude                          & Phase~(rad)                         & FF~(\%)                           & BF~($10^{-3}$)           & $\sigma$\\
      \hline
      $D_s^+\to \bar{K}^{*}(892)^0K^+  $ & 0.0(fixed)                          & $32.7 \pm 2.2 \pm 1.9$            & $4.77 \pm 0.38 \pm 0.32$ & $>10$\\
      $D_s^+\to K^{*}(892)^+K_S^0      $ & $-0.16 \pm 0.12 \pm 0.11$           & $13.9 \pm 1.7 \pm 1.3$            & $2.03 \pm 0.26 \pm 0.20$ & $>10$\\
      $D_s^+\to a_{0}(980)^+\pi^0      $ & $-0.97 \pm 0.27 \pm 0.25$           & \phantom{0}$7.7  \pm 1.7 \pm 1.8$ & $1.12 \pm 0.25 \pm 0.27$ & $6.7$\\ 
      $D_s^+\to \bar{K}^*(1410)^{0}K^+ $ & \phantom{0}$0.17 \pm 0.15 \pm 0.08$ & \phantom{0}$6.0  \pm 1.4 \pm 1.3$ & $0.88 \pm 0.21 \pm 0.19$ & $7.6$\\
      $D_s^+\to a_{0}(1710)^+\pi^0     $ & $-2.55 \pm 0.21 \pm 0.07$           & $23.6 \pm 3.4 \pm 2.0$            & $3.44 \pm 0.52 \pm 0.32$ & $>10$\\
      \hline \hline
    \end{tabular}
  \end{center}
\end{table*}
%------------------------------------------------------------------------------

The differences between the results of the nominal fit and the following
alternative fits are assigned as the systematic uncertainties for the amplitude
analysis. To estimate the systematic uncertainty related to resonances, the
masses and widths of the $\bar{K}^{*}(892)^{0}$, $K^{*}(892)^{+}$,
$a_{0}(980)^+$, and $\bar{K}^*(1410)^{0}$ are varied by their
uncertainties~\cite{PDG}. The uncertainty associated with Blatt-Weisskopf
barriers are studied by varying the radii by $\pm 1$~GeV$^{-1}$. The
uncertainty caused by background is studied by increasing or decreasing $f_s$
within its statistical uncertainty, and by varying the proportion of MC
background components according to the uncertainties of their cross section
measurement. The intermediate resonances with statistical significances less
than $5\sigma$ are included in the fit one by one and the largest difference
with respect to the baseline fit is taken as the systematic uncertainty. The
acceptance of the detector is examined by repeating the amplitude analysis fit
with different particle-identification and tracking efficiencies according to
their uncertainties. The total uncertainties are determined by adding all the
contributions in quadrature.

\begin{figure}[htp]
  \begin{center}
    \includegraphics[width=0.48\textwidth]{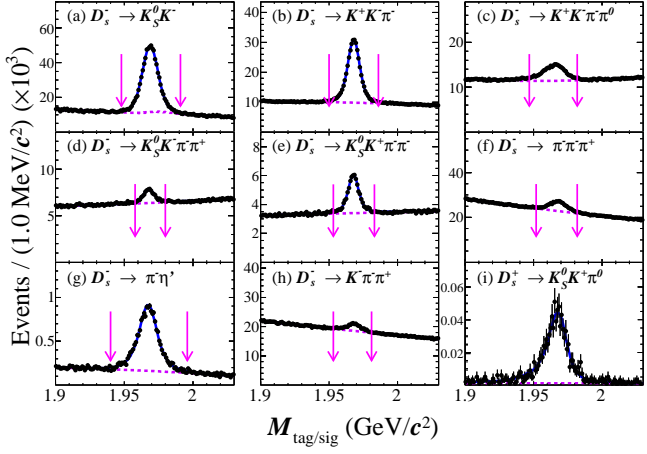}
    \caption{Fits to (a)-(h) the $M_{\rm tag}$ distributions of the ST
      candidates of various tag modes and (i) the $M_{\rm sig}$ distribution of
      the DT candidates. The data samples are represented by points with error
      bars, the total fit results by blue solid lines, and backgrounds by violet
      dashed lines. The pairs of pink arrows indicate the signal regions.
    }
    \label{fig:DT_fit}
  \end{center}
\end{figure}

%------------------------------------------------------------------------------
To measure the absolute BF of the process $D_s^+\to K_S^0K^+\pi^0$, we use the
same event selection criteria as those for the amplitude analysis, except that
the momentum of the final state $\pi^+$ originating from the signal $D_s^+$
meson is required to be larger than $0.1$~GeV/$c$ to remove soft pions from
$D^{*+}$ decays, and the best candidate strategy is changed. The best ST
candidate from the tagged $D_s^-$ is chosen using the recoiling mass closest
to the known $D_s^{*+}$ mass~\cite{PDG} per tag mode. The best DT candidate is
chosen using the average mass of the tagged $D_s^-$~($M_{\rm tag}$) and the
signal $D_s^+$~($M_{\rm sig}$) closest to the known $D_s$ mass per tag mode.
The BF of the $D_s^+\to K_S^0K^+\pi^0$ decay is determined by
\begin{eqnarray} \begin{aligned}
    \mathcal{B}(D_{s}^{+} \to K_{S}^0K^+\pi^{0})=\frac{N_{\text{sig}}^{\text{DT}}}{\sum_{\alpha}
     N_{\alpha}^{\text{ST}}\epsilon^{\text{DT}}_{\alpha,\text{sig}}/\epsilon_{\alpha}^{\text{ST}}},\, \label{eq:Bsig-gen}
\end{aligned} \end{eqnarray} 
%------------------------------------------------------------------------------
where $\alpha$ represents various tag modes. The ST yield for tag mode
$\alpha$, $N_{\alpha}^{\text{ST}}$, is obtained from fits to the $M_{\rm tag}$
distributions of the ST candidates from the data sample, as shown in
Figs.~\ref{fig:DT_fit}(a-h). The MC-simulated shape convolved with a Gaussian
function is used to model the signal shape while the background shape is
parameterized by a second-order Chebyshev function. The MC-simulated shapes of
$D^{-} \to K_{S}^{0} \pi^-$ and $D_{s}^{-} \to \eta\pi^+\pi^-\pi^-$ decays are
added to the Chebyshev functions in the fits to $D_{s}^{-} \to K_{S}^{0}K^{-}$
and $D_{s}^{-} \to \pi^{-}\eta^{\prime}$, respectively, to account for peaking
background. The DT yield, $N_{\text{sig}}^{\text{DT}}$, is determined from the
fit to the $M_{\rm sig}$ distribution of the DT candidates from the data
sample, as shown in Fig.~\ref{fig:DT_fit}(i), in which the signal shape is the
MC-simulated shape convolved with a Gaussian function and the background shape
is described by the MC-simulated background shape. The inclusive MC samples
with $D_s^+\to K_S^0K^+\pi^0$ events generated based on the amplitude analysis
are studied to determine the ST efficiencies $\epsilon_{\alpha}^{\text{ST}}$
and DT efficiencies $\epsilon^{\text{DT}}_{\alpha,\text{sig}}$. The total ST
yield of all tag modes and the DT yield are $531217\pm2235$ and $985\pm40$,
respectively. The BF of the $D_s^+\to K_S^0K^+\pi^0$ decay is determined to be
$(1.46\pm0.06_{\rm stat.}\pm0.05_{\rm syst.})\%$. The BFs for various
intermediate processes are calculated with
$\mathcal{B}_{i} = {\rm FF}_{i} \times \mathcal{B}(D_{s}^{+} \to K_{S}^0K^+\pi^{0})$
and the results are listed in Table~\ref{fit-result}.

%------------------------------------------------------------------------------

%------------------------------------------------------------------------------
The systematic uncertainties on the BF measurement from the following sources
are studied. The uncertainty in the total number of the ST $D_s^-$ mesons is
assigned to be 0.4\%, including the changes of the fit yields when varying the
signal shape, background shape, and taking into account the background
fluctuation in the fit. The uncertainty associated with the background shape
in the fit to the $M_{\rm sig}$ distribution is estimated to be 1.9\% by
replacing the nominal background shape with a second-order Chebyshev function. 
The uncertainty for the $K_{S}^{0}$ reconstruction efficiency is estimated to
be 1.5\% by using control samples of $J/\psi\to K_{S}^{0}K^{+}\pi^{-}$ and
$\phi K_{S}^{0}K^{+}\pi^{-}$ decays~\cite{prd-92-112008}.
The $K^{+}$ particle-identification and tracking efficiencies are studied with
$e^+e^-\to K^+K^-\pi^+\pi^-$ events. The data-MC differences of the $K^{+}$
particle-identification and tracking efficiencies are assigned as systematic
uncertainties, which are both 1.0\%. The systematic uncertainty of the
$\pi^{0}$ reconstruction efficiency is investigated by using a control sample
of the process $e^+e^-\to K^+K^-\pi^+\pi^-\pi^0$ and a 2.0\% systematic
uncertainty is assigned. The systematic uncertainty caused by the amplitude
analysis model is studied by varying the parameters in the amplitude analysis
fit according to the covariance matrix. The change of signal efficiency, 0.7\%,
is set as the  corresponding systematic uncertainty.

%------------------------------------------------------------------------------
In summary, this Letter presents the first amplitude analysis of the decay
$D_{s}^{+} \to K_{S}^0K^+\pi^{0}$ using 6.32 fb$^{-1}$ of $e^+e^-$
annihilation data taken at center-of-mass energies between 4.178~GeV and
4.226~GeV. The BF of $D_{s}^{+} \to K_{S}^0K^+\pi^{0}$ is determined to be
$(1.46\pm0.06_{\rm stat.}\pm 0.05_{\rm syst.})\%$, which is consistent with
the CLEO result~\cite{CLEO-BF}. The precision is improved by a factor of 2.8.

The phases and the FFs of intermediate states are listed in
Table~\ref{fit-result}. The statistical significance of
$D_{s}^{+} \to a_0(1710)^+\pi^{0}$ is found to be larger than $10\sigma$. The
mass and width of the $a_0(1710)^+$ are measured to be
($1.817 \pm 0.008_{\rm stat.} \pm 0.020_{\rm syst.}$)~GeV/$c^2$ and
($0.097 \pm 0.022_{\rm stat.} \pm 0.015_{\rm syst.}$)~GeV/$c^{2}$,
respectively. Along with the results of $D_{s}^{+} \to K_S^0K_S^0\pi^{+}$ by
BESIII~\cite{KSKSpi} and $\eta_c \to \eta \pi^+\pi^-$ by BaBar~\cite{a01710},
the isovector partner $a_0(1710)$ of the $f_0(1710)$ meson, proposed by
Refs.~\cite{Oset1, Oset2, Klempt}, is established in the process of
$D_{s}^{+} \to a_0(1710)^+\pi^{0}$ with $a_0(1710)^+\to K_S^0K^+$, whose BF is
consistent with the prediction by Ref.~\cite{Oset2}.

%------------------------------------------------------------------------------
In addition, the ratio
$\frac{\mathcal{B}(D_{s}^{+} \to \bar{K}^{*}(892)^{0}K^{+})}{\mathcal{B}(D_{s}^{+} \to \bar{K}^{0}K^{*}(892)^{+})}$
is determined to be $2.35^{+0.42}_{-0.23\text{stat.}}\pm 0.10_{\rm syst.}$.
The contribution of $D_{s}^{+} \to a_0(980)^+\pi^{0}$ is also observed.
Using $\mathcal{B}(D_{s}^{+} \to a_0(980)^+\pi^0)$~\cite{Doc-DB-682-v7}, we
determine the ratio
$\frac{\mathcal{B}(a_0(980)^+ \to \bar{K}^{0}K^{+})}{\mathcal{B}(a_0(980)^+ \to \pi^+\eta)}=(13.7 \pm 3.6_{\rm stat.} \pm 4.2_{\rm syst.})\%$.
%------------------------------------------------------------------------------

\acknowledgements
The BESIII collaboration thanks the staff of BEPCII and the IHEP computing center for their strong support. This work is supported in part by National Key R\&D Program of China under Contracts Nos. 2020YFA0406400, 2020YFA0406300; National Natural Science Foundation of China (NSFC) under Contracts Nos. 11625523, 11635010, 11735014, 11822506, 11835012, 11875054, 11935015, 11935016, 11935018, 11961141012, 12022510, 12025502, 12035009, 12035013, 12061131003, 12192260, 12192261, 12192262, 12192263, 12192264, 12192265; the Chinese Academy of Sciences (CAS) Large-Scale Scientific Facility Program; Joint Large-Scale Scientific Facility Funds of the NSFC and CAS under Contracts Nos. U2032104, U1732263, U1832207; CAS Key Research Program of Frontier Sciences under Contract No. QYZDJ-SSW-SLH040; 100 Talents Program of CAS; INPAC and Shanghai Key Laboratory for Particle Physics and Cosmology; ERC under Contract No. 758462; European Union Horizon 2020 research and innovation programme under Contract No. Marie Sklodowska-Curie grant agreement No 894790; German Research Foundation DFG under Contracts Nos. 443159800, Collaborative Research Center CRC 1044, FOR 2359, FOR 2359, GRK 214; Istituto Nazionale di Fisica Nucleare, Italy; Ministry of Development of Turkey under Contract No. DPT2006K-120470; National Science and Technology fund; Olle Engkvist Foundation under Contract No. 200-0605; STFC (United Kingdom); The Knut and Alice Wallenberg Foundation (Sweden) under Contract No. 2016.0157; The Royal Society, UK under Contracts Nos. DH140054, DH160214; The Swedish Research Council; U. S. Department of Energy under Contracts Nos. DE-FG02-05ER41374, DE-SC-0012069.

\end{document}